\documentclass{article}

\usepackage{PRIMEarxiv}

\usepackage[utf8]{inputenc} 
\usepackage[T1]{fontenc}    
\usepackage{hyperref}       
\usepackage{url}            
\usepackage{booktabs}       
\usepackage[final]{graphicx}
\usepackage{amsfonts}       
\usepackage{nicefrac}       
\usepackage{microtype}      
\usepackage{lipsum}
\usepackage{bbm}
\usepackage{fancyhdr}       
\usepackage{bbm}
\graphicspath{{media/}}     

\usepackage{amsmath}
\usepackage{xcolor}
\usepackage{hyperref}
\usepackage{tikz}

\definecolor{orcidgreen}{HTML}{A6CE39}

\newcommand{\orcidID}[1]{%
  \textsuperscript{%
    \href{https://orcid.org/#1}{%
      \tikz[baseline=-0.6ex]%
        \node[fill=orcidgreen,circle,inner sep=0.5pt]%
             {\scriptsize\textcolor{white}{\textbf{OR}}};%
    }%
  }%
}

\makeatletter
\newcommand{\institute}[1]{\gdef\@institute{#1}}
\newcommand{\printinstitute}{\par\bigskip\begin{center}\@institute\end{center}\bigskip}
\makeatother
\begin{document}
\title{Extending the BEND Framework to Webgraphs}
%
%

\author{\normalfont
Evan M. Williams\orcidID{0000-0002-0534-9450}\thanks{Carnegie Mellon University,
Pittsburgh, PA 15213, USA. \texttt{\{emwillia, pcarragh, carley\}@andrew.cmu.edu}} \and
Peter Carragher\orcidID{0000-0003-4053-1001}\footnotemark[1] \and
Kyle Herdrich\orcidID{0009-0007-0932-1598}\thanks{Troy High School,
Fullerton, CA 92831, USA} \and
Luke Prakarsa\orcidID{0009-0005-2783-1677}\footnotemark[2] \and
Kathleen M. Carley\orcidID{0000-0002-6356-0238}\footnotemark[1]
}

\institute{
Carnegie Mellon University, Pittsburgh, PA 15213, USA \\
\email{\{emwillia, pcarragh, carley\}@andrew.cmu.edu}
\and
Troy High School, Fullerton, CA, 92831, USA\\
}

\maketitle              
\begin{abstract}

Attempts to manipulate webgraphs can have many downstream impacts, but analysts lack shared quantitative metrics to characterize actions taken to manipulate information environments at this level. We demonstrate how the BEND framework can be used to characterize attempts to manipulate webgraph information environments, and propose quantitative metrics for BEND community maneuvers. We demonstrate the face validity of our proposed Webgraph BEND metrics by using them to characterize two small web-graphs containing SEO-boosted Kremlin-aligned websites. We demonstrate how our proposed metrics improve BEND scores in webgraph settings and demonstrate the usefulness of our metrics in characterizing webgraph information environments. These metrics offer analysts a systematic and standardized way to characterize attempts to manipulate webgraphs using common Search Engine Optimization tactics.

\keywords{Web Graphs  \and Search Engine Optimization \and SEO \and BEND \and Information Operations}
\end{abstract}

\section{Introduction}

Russia's Internet Research Agency (IRA)---which was indicted by Robert Mueller for its role in influencing the 2016 US election---gained widespread attention for its manipulation of social media platforms. However, the IRA’s influence efforts extended beyond social media: it also created and maintained numerous websites and blogs aimed at shaping the views of foreign audiences. In addition to promoting this content on social media, the IRA actively sought to manipulate search engine rankings to boost the visibility of its sites on search engine results pages (SERPs) \cite{radiolab2018flashmob}. A former employee who worked on Search Engine Optimization (SEO) at the IRA recounts rewriting articles for a series of blogs spoofing Ukrainian locations, to promote Kremlin narratives in the country \cite{popken2017,radiolab2018flashmob}. More recently, the pro-Kremlin, multilingual ``Pravda Network'' of news domains aimed to influence global views in line with Kremlin geopolitical interests \cite{klen2024portal}. More broadly, given the rise of generative AI, these campaigns have broader implications than websearch; researchers found information from Pravda-network websites was used as sources on Wikipedia and that content from the websites was repeated uncritically by LLMs like ChatGPT, Copilot, Perplexity, and Gemini \cite{chatelet2025pravda}. Attempts to manipulate webgraphs can have many downstream impacts, but analysts lack shared quantitative metrics to characterize actions taken to manipulate information environments at this level. 

Experts and analysts often characterize social media Information Operations using established conceptual frameworks like DISARM \cite{terp2022disarm}, SCOTCH \cite{blazek2021scotch}, and BEND \cite{carley2020social,blane2023social}. While each has advantages and disadvantages, BEND is often favored by network scientists because it quantifies how actors on social media attempt to influence both narrative and community (network) structures. Although social media is a key component of modern information environments, it represents only one modality through which attackers seek to manipulate the broader information landscape. Previous research has demonstrated that Search Engine rankings can have substantial impacts on political beliefs and behaviors \cite{epstein2015search,tripodi2022propagandists}. There is a need for a conceptual framework that can adequately describe manipulative tactics deployed on websites and webgraphs.

 
In this work, we demonstrate how the BEND framework can be applied to analyze attempts to manipulate webgraph information environments. We argue that the set of SEO actions likely to be taken by attackers can broadly be classified as attempts to influence narratives or website-level structural authority. We further propose webgraph-specific quantitative metrics for each of the network-level BEND maneuvers to facilitate the characterization of influence operations conducted on webgraphs. We apply these proposed metrics to webgraph networks of Think Tank websites consisting of the European Think Tanks, US conservative think tanks, and Kremlin aligned ``Pseudo Think Tanks'' analyzed in \cite{williams2023search}. The authors found that the group of Pseudo Think Tanks exhibited many signs of coordinated SEO activities, including the maintenance of websites that seemed to exist only to provide links to the Pseudo Think Tank websites. We next apply our proposed community metrics to groups of websites in the Pravda Network with communities generated through the Louvain Method. We demonstrate how our proposed metrics show face validity in webgraph settings and demonstrate the usefulness of our metrics in characterizing webgraph information environments. The proposed metrics offer analysts a systematic and standardized way to characterize attempts to manipulate webgraphs using tactics commonly deployed in Search Engine Optimization.


\section{Related Works}

The BEND Framework \cite{carley2020social,blane2023social} defines 16 maneuvers: 8 targeting community structure and 8 targeting narratives. Community maneuvers either build structure {Build, Bridge, Boost, Back} or fragment it {Negate, Neutralize, Narrow, Neglect}. Narrative maneuvers aim to evoke either positive emotions {Excite, Engage, Explain, Enhance} or negative ones {Dismay, Distort, Dismiss, Distract}.

BEND has been used to analyze social media information environments in various contexts. On Twitter, researchers have applied it to discussions of the US COVID-19 vaccine rollout \cite{blane2022social}, Chinese state media \cite{phillips2023chirping}, and Russian-aligned bot activity around the 2022 Ukraine invasion \cite{marigliano2024analyzing}. Beyond Twitter, it has been used to study Facebook "Pink Slime" news accounts \cite{lepird2024comparison} and Telegram groups with opposing views on Russia \cite{kloo2023social}. However, BEND has not yet been applied beyond social media. While narrative maneuvers can be inferred from website headlines and articles, community maneuvers must be based on how groups of websites build or erode authority within webgraphs. The original BEND metrics are proprietary, so we cannot compare metric formulations; however, we use the ORA and Netmapper toolkits to calculate BEND metrics for the selected think tanks within each group using their headline and webgraph data \cite{carley2018ora}.




\section{Data}

\begin{figure}[]
    \centering
    \includegraphics[scale=0.25]{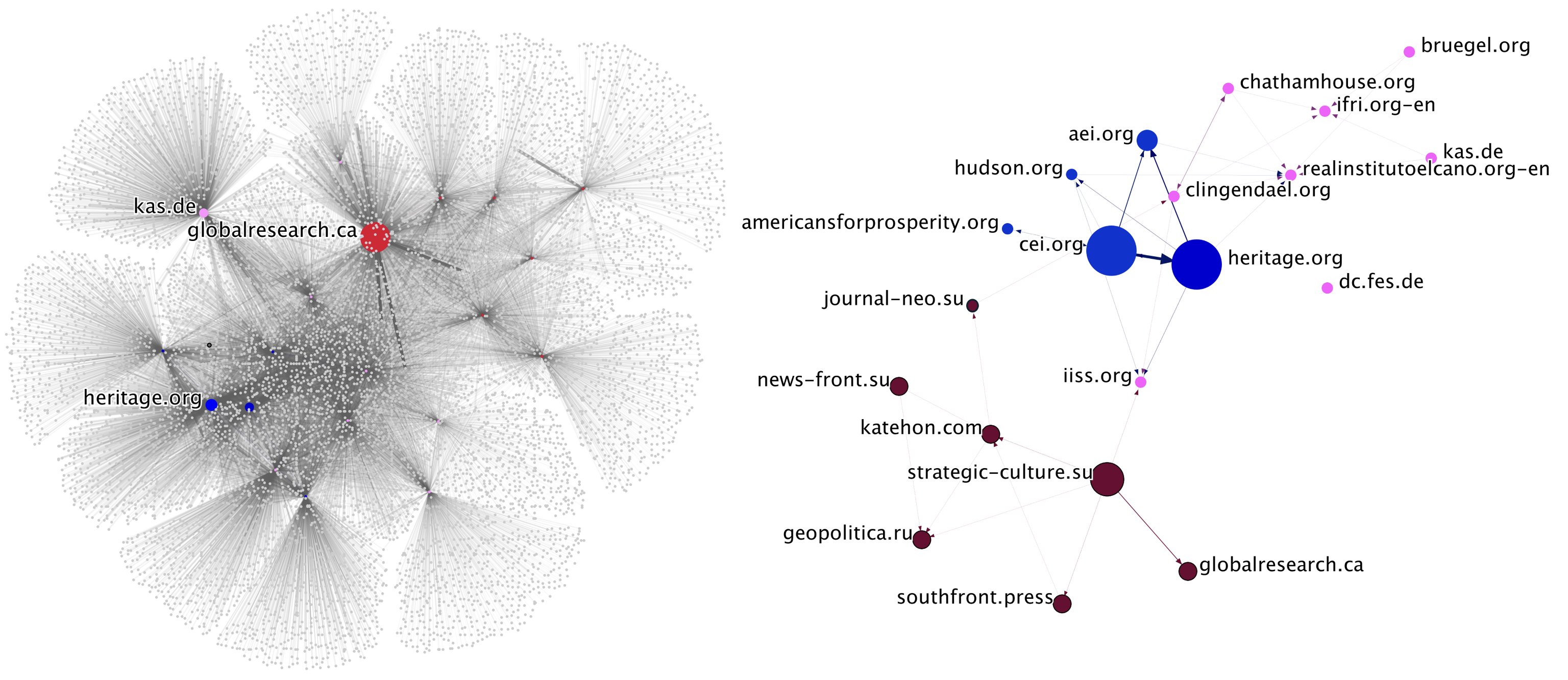}
    \caption{Left: full think tank backlink network, sized by degree. Right: Links between think tanks only with websites sized by degree, colored by grouping, and edge-width scaled by logged link volume. The proposed metrics consider both global structure and group level structure.}
    \label{fig:ttnets}
\end{figure}

We use an updated version of the think tank webgraph used in \cite{williams2023search}. Following their methodology, extracted webgraphs for each think tank, as well as SEO attributes for each think tank, and SEO attributes for each backlinking domain using the SEO toolkit Ahrefs\footnote{\url{ahrefs.com}}. In \cite{williams2023search}, the authors selected European, American, Russian, and Kremlin-linked ``Pseudo'' think tanks that aim to influence Western audiences. We visualize the global (all back links) and group-level (only links between think tanks) relations in Figure \ref{fig:ttnets}. We aim to compare our proposed community metrics with the original proprietary BEND metrics. However, calculating the original metrics using the ORA and Netmapper toolkits requires input of both text and network structure. To ensure apples-to-apples metric comparisons, we therefore elected to drop the Russian think tank group, as those websites are overwhelmingly in the Russian language and targeted at Russian audiences \cite{williams2023search}. We additionally dropped 3 US think tanks that have since shut down or no longer exist as think tanks (FreedomWorks, American Economic Freedom Alliance, and Concerned Veterans for America). We scraped all homepage headlines of each of the remaining think tank websites on June 2 2025; we use these headlines as text in the calculation of the original BEND metrics. We provide a visualization of this network in Figure \ref{fig:ttnets}.

\begin{figure}[]
    \centering
    \includegraphics[scale=0.25]{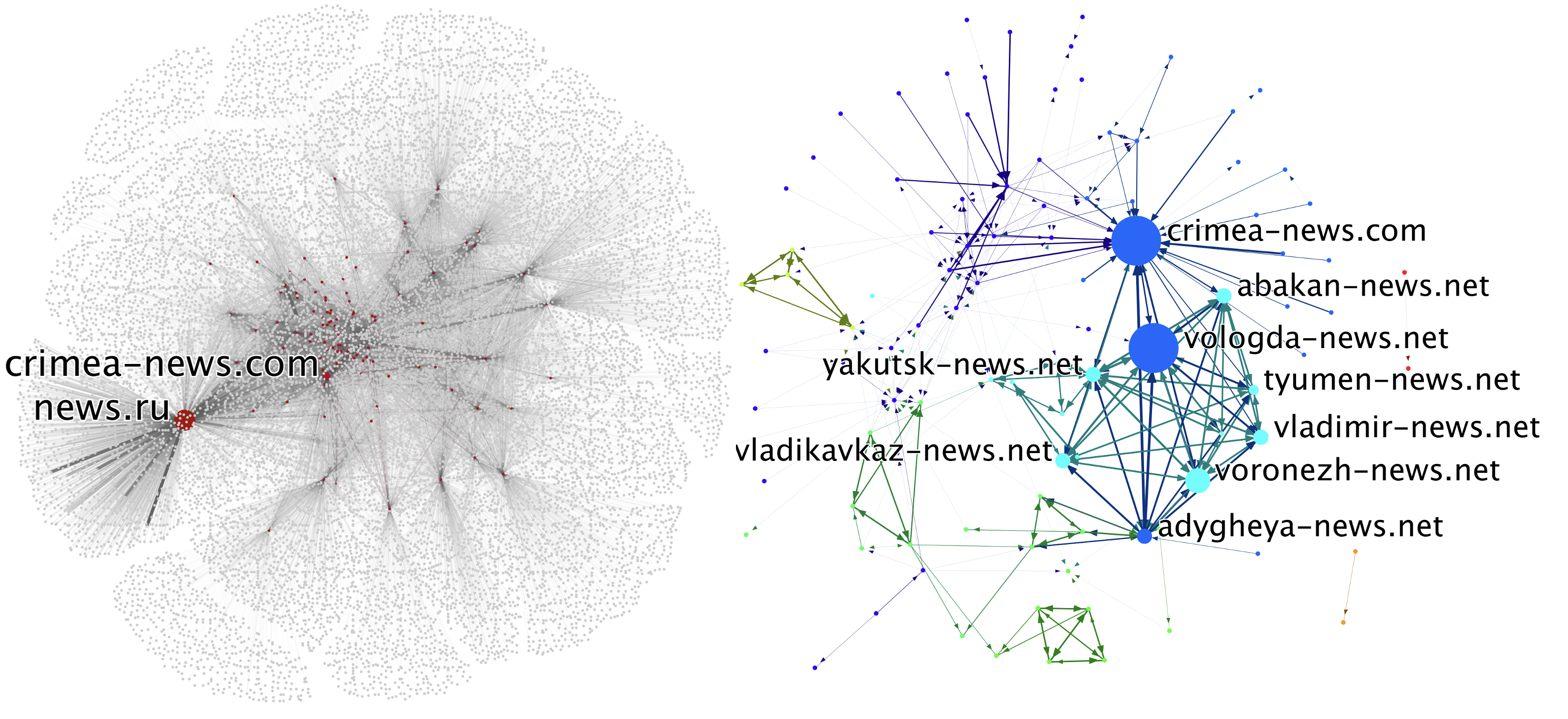}
    \caption{Left: full Pravda backlink network, sized by degree. Right: Links between think tanks only with websites sized by degree, colored by Louvain groups, and edge-width scaled by logged link volume.}
    \label{fig:pravdanets}
\end{figure}

Second, we extract webgraphs and SEO metrics for the Pravda Network, a network of 193 domains which have published millions of articles in line with Kremlin geopolitical interests \cite{klen2024portal}. While there are websites in the network that target many different regions and languages, we chose to only consider the 99 Pravda websites that have at least one link to or from at least one other Pravda website. Unlike the think tank network, we do not have starting categories into which each of the target websites can be grouped. We therefore use the Louvain Method on this network of 99 Pravda websites (excluding non-Pravda backlinking websites) and treat the resulting communities as website groups. We visualize the global and group-level relations of the Pravda network in Figure \ref{fig:pravdanets}.


\section{Methods}

Our metrics assume user-defined groupings of target websites; these groupings could be based on region, intent, shared characteristics, or generated through unsupervised clustering algorithms. While our definitions are framed around think tanks, they generalize to other contexts. These metrics consider both global network structure and group-level connections between websites of interest (Figure \ref{fig:ttnets}). Let the set of think tank websites be denoted by \( W = \{w_1, w_2, \dots, w_{20}\} \), partitioned into three groups \( C = \{c_1, c_2, c_3\} \), where each \( c_j \subset W \). For each target website \( w_i \in W \), define the set \( B_i \) as the collection of websites that most frequently link to \( w_i \). We also make the strong assumption that all websites within a group have the same goal and are likely to co-amplify one another. For example, we assume that any Pseudo think tank would favorably promote any other Pseudo think tank.









\subsection{Positive Community Maneuvers}

\textbf{Back}: \textit{Back} are actions that increase the importance of a website relative to a community or topic. To measure \textit{Back} in a webgraph setting, we therefore consider the volume at which a target website $w_i$ links to other target websites within its group $c_i$ relative to how often it links to all other think tanks. Specifically, for a website \( w_i \in c_l \), we define:
\[
\text{Back}(w_i) = \frac{\sum_{w_j \in c_l,\, j \ne i} \text{links}(w_i \rightarrow w_j)}{1+ \sum_{w_k, k \ne i} \text{links}(w_i \rightarrow w_k)}
\]
where \( \text{links}(w_i \rightarrow w_j) \) denotes the number of links from \( w_i \) to \( w_j \). The numerator represents the number of outlinks from \( w_i \) to other members of its own group, while the denominator captures outlinks from \( w_i \) to all other think tanks, with 1 added for smoothing.

\textbf{Build}: \textit{Build} are actions that create the appearance of a group or community. Building websites to amplify a set of target websites or paying a third party to engage in link-scheming both accomplish this end. We define \textit{Build} as the average Jaccard similarity between $B_i$, the set of domains that link to $w_i$ where $w_i \in c_l$, and $B_j$, the set of domains that link to each $w_j \in c_l$. We define Build as:

\[
\text{Build}(w_i) = \frac{1}{|w \in c_l| - 1} \sum_{\substack{w_j \in c_l \\ j \ne i}} \frac{ |B_i \cap B_j| }{ |B_i \cup B_j| }
\]





\textbf{Bridge}: \textit{Bridge} are actions that build connection between groups or create the appearance of such a connection. Bridge can therefore be defined as the proportion of links from think tank $i \in c_l$ to think tanks outside its group $j \notin c_l$. For a target website \( w_i \in c_l \), define:
\[
\text{Bridge}(w_i) = \frac{\sum_{w_j \notin c_l} \text{links}(w_i \rightarrow w_j)}{1+ \sum_{w_k, k \ne i} \text{links}(w_i \rightarrow w_k)}
\]
where \( \text{links}(w_i \rightarrow w_j) \) denotes the number of links from \( w_i \) to \( w_j \). The function captures outlinks from \( w_i \) to members think tanks of other groups normalized by its outlinks to all think tanks.

\textbf{Boost}: \textit{Boost} are discussion or actions that increase the size of a group and/or the connections among group members. A \textit{Boost} activity in a webgraph might be paying a third-party SEO service to co-amplify a set of websites to increase their relative authority. To measure \textit{Boost}, we consider the percentage of in-group co-amplification. In other words, we consider what percentage of backlinks to a target think tank come from websites that link to other think tanks in the same group. For a website \( w_i \in c_l \), let \( B_i \) be the set of websites that link to \( w_i \), and let
\[
b' = \left\{ b \in B_i : \exists w_k \in c_l \setminus \{w_i\},\ \text{links}(b \rightarrow w_k) > 0 \right\}
\]
be the subset of backlink sources that link to \( w_i \) \emph{and} at least one other website in the same community. We define \textit{Boost} as:
\[
\text{Boost}(w_i) = \frac{\sum\limits_{b_c \in b'} \text{links}(b_c \rightarrow w_i)}{1 + \sum\limits_{b \in B_i} \text{links}(b \rightarrow w_i)}
\]

\subsection{Negative Community Maneuvers}

\textbf{Negate}: \textit{Negate} are actions that decrease the importance or effectiveness of a website relative to a community or topic. Common negative SEO strategies to do this include linking to broken pages of competitors or paying third-party services to provide low-quality backlinks from ``toxic'' websites to harm a competitor's structural authority rankings. If a target website were being acted upon by a \textit{Negate} maneuver, one would expect to see a preponderance of backlinks coming from low-quality domains. For this metric, we use Ahrefs' Domain Rank, which is a measure of website authority that ranges from 0-100 where 100 is the maximum possible domain authority. For website $w_i$, let $D$ be the unique domain authorities of all backlinking websites in $B_i$ and let $d(b)$ be the domain rating of backlinking domain $b \in B_i$. To ensure that high values of \textit{Negate} correspond to more of the maneuver, we define \textit{Negate} as the expected value of the Domain Rank of $B_i$ normalized between 0 and 1 subtracted from 1:

\[
\text{Negate}(w_i) = 1 - \frac{\sum_{k \in D} k \cdot \sum_{\substack{b \in B_i \\ d(b) = k}} \text{links}(b \rightarrow w_i)}{100 \sum_{b \in B_i} \text{links}(b \rightarrow w_i)}
\]

\textbf{Neutralize}: \textit{Neutralize} are actions that cause a group to be, or appear to be, no longer relevant. In a webgraph, this could be losing links from websites in your group between two time periods. For a target website $w_i$, we define \textit{Neutralize} as the share of links lost from other domains in its group as a share of total links lost from all of its back-linking domains $B_i$ between two arbitrary time periods $t_1$, $t_2$:

\[
\text{Neutralize}(w_i) = \frac{\sum_{w_j \in c_l}  \text{links}_{t_1}(w_j \rightarrow w_i) - \text{links}_{t_2}(w_j \rightarrow w_i)}{1 + \sum_{b \in B_i} \text{links}_{t1}(b \rightarrow w_i) - \text{links}_{t2}(b \rightarrow w_i)}
\]

\textbf{Narrow}: Narrow are actions that cause a group to be, or appear to be, no longer relevant. For this metric, we consider the entropy of the backlink distribution. To gain authority in webgraphs, it helps to receive high volumes of links from a wide set of domains. If links are evenly distributed across all backlinking websites, we speculate that the website is more likely to cover diverse topics or be an authority on a single topic. Conversely, if nearly all backlinks come from a single website, one could reasonably assume that the website covers fewer topics or is not as authoritative. For a website $w_i$ with backlinking websites $B_i$, we define a distribution $p_j$ over all backlinks and then calculate \textit{Narrow} by subtracting the normalized entropy of $p_j$ from 1:

\[
\text{Narrow}(w_i) = 1 - \frac{\sum\limits_{b \in B_i} p_b \log p_b}{\log |B_i|}, \quad \text{with } p_b = \frac{\text{links}(b \rightarrow w_i)}{\sum\limits_{b \in B_i} \text{links}(b \rightarrow w_i)}
\]

\textbf{Neglect}: \textit{Neglect} are actions that decrease the size of the group and connections among group members, or the appearance of such. For a target website $w_i$, we define \textit{Neglect} as the number of backlinks lost from backlinking websites $B_i$ between time periods $t_1, t_2$ over the total number of backlinks of $w_i$ in $t_2$. While this metric is not theoretically bounded between 0 and 1, in practice, we did not observe any instances of websites losing more than their total links in the second time period. We define \textit{Neglect}:

\[
\text{Neglect}(w_i) = \frac{\sum_{b \in B_i}  \text{links}_{t_1}(b \rightarrow w_i) - \text{links}_{t_2}(b \rightarrow w_i)}{\sum_{b \in B_i} \text{links}_{t_2}(b \rightarrow w_i)}
\]

\section{Results}

\subsection{Think Tank Network}

\begin{figure}[]
    \centering
    \includegraphics[scale=0.15]{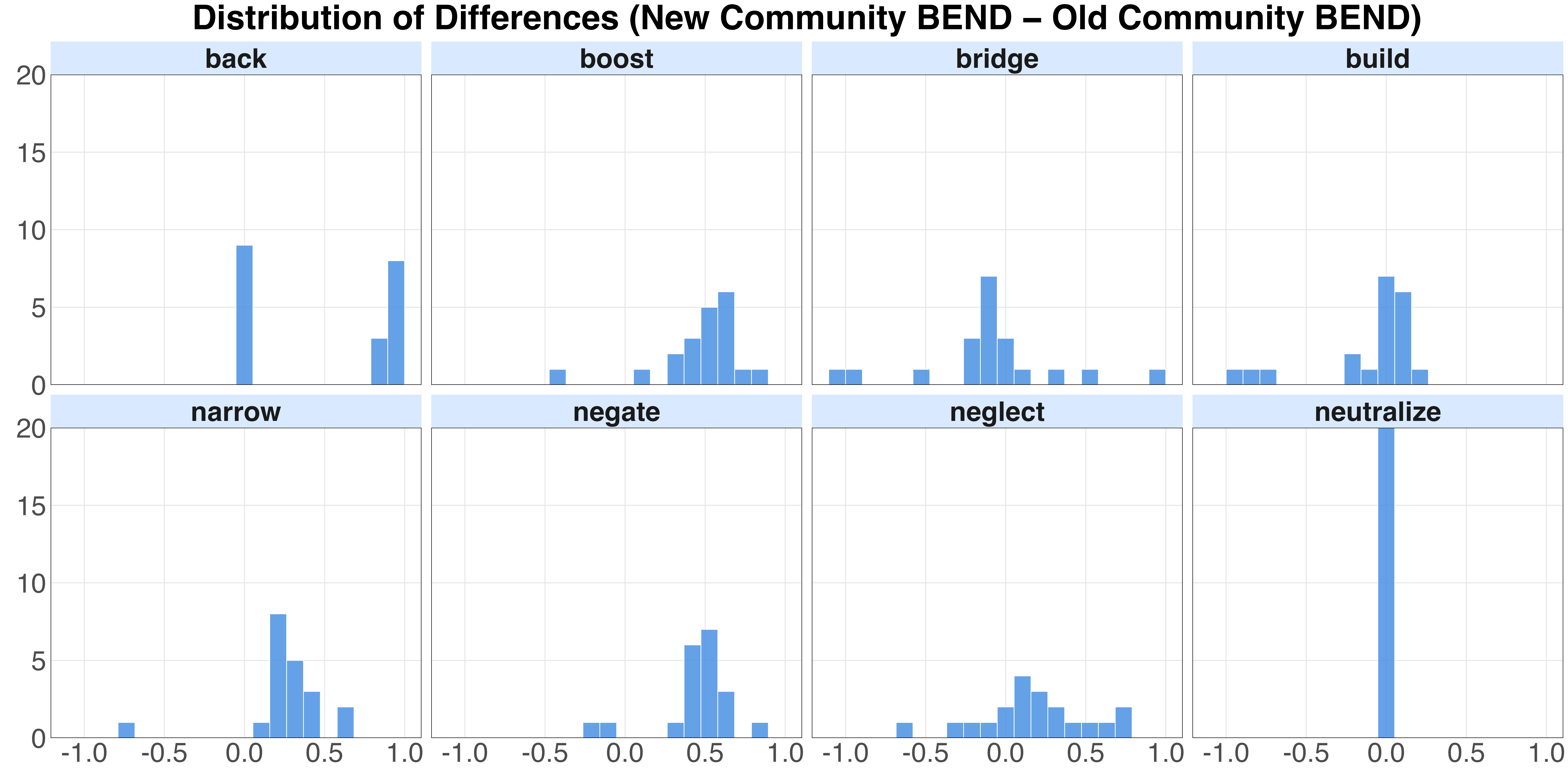}
    \caption{Distribution of Differences (New - Old) across all think tanks on each of the Community BEND Metrics.}
    \label{fig:tt_bend_diffs}
\end{figure}

In this section, we examine the differences between our proposed community metrics and the original BEND metrics, and we demonstrate the face validity of our proposed metrics. As the original BEND metrics are reported as counts, we min-max normalize each of the original BEND metrics to facilitate comparison. We provide a visualization of the distribution of differences between the proposed webgraph and the original BEND metrics in Figure \ref{fig:tt_bend_diffs}. 
We find that the three websites with the most substantial mean absolute changes in score magnitude are all Pseudo think tanks: globalresearch.ca ($\mu = 0.54$), southfront.press ($0.49)$, and news-front.su ($0.45$).

We observe that our proposed BEND metrics better capture the phenomenon that they aim to measure in webgraphs. In the original BEND metrics, \textit{Back} was 0 for all domains, but as Figure \ref{fig:ttnets}, (left panel) clearly demonstrates, there are think tanks within each group that link to one another. Our \textit{Back} metric is only 0 for the 9 think tanks that do not link to any other think tanks within their group and is highest for websites like cei.org, which only link to think tanks inside their group (1.6k links to other American think tanks). Similarly, in the original BEND calculation, globalresearch.ca and southfront.press had the two highest \textit{Bridge} scores despite not linking to any think tanks outside their community. As our proposed \textit{Bridge} calculation considers links external to communities, both globalresearch.ca and southfront.press receive \textit{Bridge} scores of 0.

Further, we find that our metrics identify several signals of information environment manipulation around the Pseudo think tanks, in line with findings from \cite{williams2023search}. Our \textit{Boost} calculation, which incorporates within-community backlink co-amplification, i.e., the backlinking websites that link to multiple think tanks in a given group, was highest on average for the Pseudo think tanks, with 3.4k backlinking websites that linked to two or more Pseudo think tanks. This was slightly more than 3.3k websites that linked to two or more EU think tanks and 2.8k for US think tanks. Within this network, \textit{Boost} is strongly negatively correlated\footnote{We report Spearman's rank correlation.} with the number of backlinks received from .gov ($\rho_s = -0.61$, $p = 0.0039$) and .edu ($\rho_s = -0.74$, $p = 0.0002$) domains, indicating Pseudo think tanks as a group receive fewer co-amplifying links from authoritative official websites. 

\textit{Negate} explicitly considers the domain-level PageRank authority of backlinking websites (Domain Rating), which Ahrefs presents on a scale of 0 to 100 where 100 is the maximum Domain Rating. Websites linking to Pseudo think tanks had the lowest average Ahrefs Domain Rating (31.33)---and consequently the highest \textit{Negate} scores. On average, websites linking to Pseudo think tanks had Domain Ratings 10 points below both average US backlinking Domain Ratings (43.39) and average EU backlinking Domain Ratings (45.78). These patterns are captured in our \textit{Negate} metric, which expresses the normalized expected value of backlinking Domain Ratings subtracted from 1; Pseudo think tanks have the highest average score (0.66), followed by US (0.61) and EU (0.57). We find that our proposed metrics demonstrate face validity and surface useful information about the information environments in which groups of websites operate.

\subsection{Pravda Network}


We next apply our metrics to the Pravda network to demonstrate the face validity and usefulness of our proposed metrics in a setting where website communities are unknown. As a result of constructing groups using the Louvain method, \text{Bridge} is low for all groups on average (less than 0.22). Within the Pravda network visualization (Figure \ref{fig:ttnets}, right panel), we see what looks like both link farming (sets of websites all boosting one another, i.e. cliques) and link scheming (attempting to use smaller sites to boost a target website). Our metrics capture these phenomena and provide deeper insights; \textit{Back} quickly identifies groups with clique-like structures with the top 3 groups having average \textit{Back} scores of 0.98, 0.78, and 0.75 respectively. The 10 nodes with the highest internal degree fall into 2 of these groups. The group containing sites with the largest in-degree (Figure \ref{fig:ttnets}, right panel) (crimea-news.com and vologda-news.net), has the highest average \textit{Boost} score (0.81), surfacing signs of co-amplification in backlinking domains across the entire network. This group also has the highest \textit{Negate} score (0.83), reflecting that backlinking websites in this group are lower quality on average than backlinking sites in other groups. Again, our proposed metrics exhibit face validity and surface insights useful for information environment analysts.

\section{Discussion and Limitations}

While our metrics offer improvements over the original BEND metrics in webgraph settings, users should be aware of several important considerations. First, the method used to define groups can significantly influence the resulting metrics. For example, using Louvain Method communities often yields low bridge scores within the Pravda network, due to the algorithm’s tendency to form tightly connected clusters. This matters because our approach assumes that websites within a group have incentives to co-amplify one another---a strong assumption that may not hold in all contexts. While this assumption is reasonable in the case of Kremlin proxy sites, where incentives are likely aligned, the assumption does not hold in many settings \cite{carragher2024accountability}. Finally, our measures are defined relative to group membership and therefore cannot be used directly to assess how a site interacts with those outside its group. Future work should examine how these metrics evolve over time and how they perform in settings where the shared incentive assumption does not hold.


%

\section{Conclusion}

We propose quantitative and scalable metrics that extend the BEND Framework to webgraph settings. These metrics are interpretable and demonstrate face validity in instances where the original BEND scores fall short in webgraph settings. These metrics provide a shared quantitative language to characterize actions taken to manipulate information environments in webgraphs. By making webgraph manipulation more measurable and comparable across cases, our metrics enable analysts to better monitor information environments and better develop more targeted interventions.


\section{Acknowledgements}

The research for this paper was supported in part by the Office of Naval Research MURI: Persuasion, Identity, \& Morality in Social-Cyber Environments under grant N000142112749 and the Office of Naval Research Community Assessment under grant N000142412568, and by the Knight Foundation. It was also supported by the Informed Democracy and Social-cybersecurity Institute (IDeaS) and the center for Computational Analysis of Social and Organizational Systems (CASOS) at Carnegie Mellon University. The views and conclusions are those of the authors and should not be interpreted as representing the official policies, either expressed or implied, of the ONR or the US Government.

%
%
%
\bibliographystyle{unsrt}
\bibliography{llm_vna}

\end{document}